\documentclass[aps,amsfonts,showpacs]{revtex4-1}
\usepackage{amsmath}
\usepackage{graphicx}

\usepackage{natbib}
\usepackage{psfrag}
\usepackage{amssymb}
\usepackage{color}
\usepackage{hyperref}

\begin{document}

\def\be{\begin{equation}}
\def\ee{\end{equation}}
\def\bea{\begin{eqnarray}}
\def\eea{\end{eqnarray}}
\def\bml{\begin{mathletters}}
\def\eml{\end{mathletters}}
\def\b{\bullet}
\def\eqn#1{(~\ref{eq:#1}~)}
\def\no{\nonumber}
\def\av#1{{\langle  #1 \rangle}}
\def\m{{\rm{min}}}
\def\M{{\rm{max}}}
\newcommand{\ds}{\displaystyle}
\newcommand{\tc}{\textcolor}
\newcommand{\TODO}[2][To do: ]{{\textcolor{red}{\textbf{#1#2}}}}

\title{Two-point correlation function of an exclusion process with hole-dependent rates}   
\author{Priyanka$^1$, Arvind Ayyer$^2$, Kavita Jain$^1$}
\affiliation{$^{1}$Theoretical Sciences Unit, Jawaharlal Nehru Centre for Advanced Scientific Research, Jakkur P.O., Bangalore 560064, India and $^2$ Department of Mathematics, Indian Institute of Science, Bangalore  560012, India.}
\widetext
\date{\today}
\begin{abstract}
We consider an exclusion process on a ring in which a
particle hops to an empty neighbouring site with a rate that 
depends on the number of vacancies $n$ in front of it. In the steady
state, using the well known mapping of this model to the zero range
process, we write down an exact formula for the 
partition function and the particle-particle
correlation function in the canonical ensemble. In the thermodynamic
limit, we find a simple 
analytical expression for the generating function of the correlation function. 
This result is applied to the hop rate $u(n)=1+(b/n)$ for 
which a phase transition between high-density laminar phase and
low-density jammed phase occurs for $b > 2$. For these rates, we find
that at the critical density, the 
correlation function decays algebraically with a continuously varying
exponent $b-2$. We 
also calculate the two-point correlation function above the critical
density, and find that the correlation length diverges with a 
critical exponent $\nu=1/(b-2)$ for $b < 3$ and $1$ for $b > 3$. These
results are compared with those obtained using an exact series
expansion for finite systems.  
\end{abstract}
\maketitle

\section{Introduction}

Nonequilibrium steady states, which are characterised by a lack of
detailed balance, have the important property that they can exhibit
phase transitions even in one dimension \cite{Privman:2005}. The condensation
transition \cite{Evans:2005} is an important example of such a 
transition in which 
particles are distributed homogeneously over the system at low
densities, but above a critical density, a macroscopic number of 
particles form a cluster. 
This transition has been
studied using several different models including aggregation-diffusion
models and zero range processes (ZRP) for homogeneous systems
\cite{Majumdar:1998,Evans:2005} and for systems with quenched disorder
\cite{Jain:2001,Jain:2003}. The ZRP, in which a particle
hops to a neighbouring site with a rate that depends only on the 
properties of the departure site, has the attractive
feature that its steady state distribution can be found  exactly 
\cite{Spitzer:1970}. This 
model has been generalised in various directions in recent years 
\cite{Angel:2007,Grosskinsky:2008,Evans:2014}, and has been used to
model clustering phenomena in traffic flow 
\cite{Kaupuzs:2005}, granular gases \cite{Torok:2005} and networks
\cite{Mohanty:2008}, avalanche dynamics in sandpiles \cite{Jain:2005}, 
slow dynamics in glasses \cite{Ryabov:2014}, and to understand
phase separation in nonequilibrium systems \cite{Kafri:2002}.

The jamming transition is an avatar of the condensation transition, and
has been studied in an exclusion process (EP) in which a hard core 
particle hops to an empty nearest neighbour with a rate that depends
on the vacant sites in front of it. An exclusion process with hole-dependent rate mimics traffic flow on a highway where a driver increases (decreases) its speed if the headway in front of it is large (small) \cite{Chowdhury:2000,Kanai:2007}. In such traffic models, the hop rate is an increasing function of the vacancies with appropriate lower and upper bounds on the speed of the car. 
However it has been shown that if the hop rates decay sufficiently slowly with the increasing number of  vacancies, as the total density of the system
is decreased in a closed one-dimensional system, a 
transition occurs from a laminar phase with typical inter-particle
spacing of order unity to a jammed phase in which a macroscopic
headway forms in front of a particle \cite{Evans:2005}. 
Since the EP with
hole-dependent rates can be exactly mapped to a ZRP, a lot is
known about its steady state properties; however a basic question
regarding the spatial correlation functions in the EP
has not been addressed in previous studies. In this article, we are
interested in calculating the 
particle-particle correlation function in the steady state of this
model in the canonical ensemble.

Analytical formulae for the two-point correlation functions are hard to come
by. For the one-dimensional totally asymmetric simple exclusion
process (TASEP) on a ring, which is a special case
of the exclusion model 
studied here, this is trivial because all configurations are equally
likely in the steady state. A nontrivial exact formula for the TASEP
with open boundaries (entrance rate $\alpha$ and exit rate $\beta$)
was given in \cite{Derrida:1993a} for arbitrary system size.   
Although the exact formula was determined in the latter case, their
limiting behaviour has not been calculated to the best of our
knowledge, especially at the critical phase line $\alpha = \beta <
\frac{1}{2}$. Recently the particle-particle correlation function for
the EP with hole-dependent rates  
was calculated in the laminar phase for certain special choices of
hopping rates in the grand canonical ensemble \cite{Basu:2010a}.  
Here we study the same model in the canonical ensemble, and find a
simple analytical formula for the generating function of the two-point
correlation function with arbitrary hop rates in the thermodynamic
limit. This result is applied to the hop rate $u(n)=1+(b/n)$ which 
decays to a nonzero constant with the number of holes $n$ in front of
the particle. For this choice, a jamming transition occurs when $b >
2$ at a critical density $\rho_c=(b-2)/(b-1)$ \cite{Evans:2005}, and
here we calculate the two-point correlation function at the critical point
and in the laminar phase.  

The plan of the article is as follows: in Sec.~\ref{model}, we 
define the model and briefly review its steady state properties. In
Sec.~\ref{partfn}, we focus on the canonical partition function and
give a formula for it in terms of integer partitions. We then turn to
a calculation of the steady state  
particle-particle correlation function in the canonical ensemble in
Sec.~\ref{twocorrfn} and obtain
an exact expression for it for any system size. We then find an exact
expression for the generating 
function of the correlation function in the thermodynamic limit. In
Sec.~\ref{specific}, for $u(n)=1+(b/n)$, we show that at the critical 
density, the correlation function decays as a power law with continuously
varying exponent. The behaviour of the correlation function in the
laminar phase is also studied. We finally conclude with a summary of
our results and  discussion in Sec.~\ref{disc}.

\section{Model and its steady state}
\label{model}

We consider an exclusion process defined on a ring with $L$ lattice
sites and $N$ 
particles in which each site can be occupied by at most one
particle. A particle hops to its  
right empty neighbour with a rate $u(n)$ where $n$ is the number of
holes in front of the particle. This EP can be mapped
to a one-dimensional ZRP with periodic boundary
conditions in which a site can support any number
of particles and a particle hops to its left neighbour with a rate
$u(n)$ where $n$ now is the number of particles at the departure
site \cite{Evans:2005}. 
As a result of
this mapping, the density $\rho$ in the EP with $L$ sites and $N$
particles is related to the density $\varrho$ in ZRP with $l(=N)$ sites
and $n(=L-N)$ particles as $\varrho={(1-\rho)}/{\rho}$. 

As we will be exploiting the connection of EP to ZRP in the following
sections, below we briefly review the steady state properties of the
ZRP and refer the reader to \cite{Evans:2005} for details.  
The ZRP has the important property that the single site weights
factorise. More precisely, the distribution of a 
configuration $C \equiv \{m_1,...,m_l\}$, where $m_i$ is the number of
particles at the $i$th site and $\sum_{i} m_i=n$, is given by
\be
P(C)= {{\tilde Z}_{l,n}}^{-1} ~\prod_{i=1}^l f(m_i)
\label{measure}
\ee
where the single site weight $f(m)$ is
\be
f(m)=(1-\delta_{m,0}) \prod_{i=1}^m \frac{1}{u(i)} + \delta_{m,0}
\ee
and ${\tilde Z}_{l,n}$ is the  partition function of the
ZRP in the canonical ensemble given by 
\be
{\tilde Z}_{l,n}= \sum_{0 \leq m_1, \dots, m_l \leq n} \;\;
\prod_{i=1}^l f(m_i) ~\delta(\sum_{k=1}^l m_k-n)
\label{product}
\ee
In general, it is difficult to obtain results in the canonical
ensemble (see however  \cite{Evans:2006,Kanai:2007,Chleboun:2010}), but the
grand canonical partition function ${\tilde {\cal Z}}_l$ can be
readily obtained. Using (\ref{product}), we can write 
\be
{\tilde {\cal Z}}_l(z) = \sum_{n=0}^\infty {\tilde Z}_{l,n} z^n = g^l(z)
\ee
where $g(z)$ is the generating function of $f(m)$ defined as 
\be
g(z)=\sum_{m=0}^\infty z^m f(m)
\label{gzdefn}
\ee
with a radius of convergence $z^*$.
The number distribution at a site is given by $p(m)=z^m f(m)/g(z)$
where the fugacity $z (\leq z^*)$ is determined by
\be
\varrho= \frac{1}{\rho}-1=\frac{z}{l}~\frac{\partial \ln {\tilde {\cal
      Z}}_l(z)}{\partial z}=z \frac{\partial \ln g(z)}{\partial z}
\label{fugacity}
\ee
The fugacity $z$ is an increasing function of the ZRP density 
$\varrho$. But as it is bounded above, it may happen that $z$ reaches
its maximum value at a finite critical density $\varrho_c$. In such a
case, the distribution $p(m)=z^{*m} f(m)/g(z^*)$ for all $\varrho \geq
\varrho_c$. But this implies that the average density in the system is
$l^{-1} \sum_{m=1}^\infty m ~p(m)=\varrho_c$. The excess mass
$\varrho-\varrho_c$ is then said to be condensed into a single cluster. 

In Sec.~\ref{specific}, we will consider hop rates for which the
jamming transition occurs. We will work with 
\be
 u(n)=1+\frac{b}{n}~,~b > 0
\label{hoprate}
\ee
where $b$ is a constant and $n$ is number of vacant sites in front of
a particle in the EP picture. When $b=0$, we arrive at the TASEP on a
ring in which a particle hops to the right empty site irrespective of
the vacancies in front of it (see Appendix~\ref{app_b0} also).  
For the above choice of hop rates, the weight $f(n)$ is given by 
\be
f(n)=\frac{n !}{(b + 1)_n}
\label{fn}
\ee
where $(a)_n=a (a+1) ... (a+n-1)$ is the Pochhammer symbol or rising
factorial. Its generating function $g(z)$ is writeable as 
\be
g(z)={_2}F_1(1,1;1+b;z)
\label{gz}
\ee
where the Gauss hypergeometric function is defined as \cite{Abramowitz:1964}
\bea
{_2}F_1(a,b;c;z) = \sum_{n=0}^\infty \frac{(a)_n (b)_n}{(c)_n}
~\frac{z^n}{n!} 
\eea
It is easy to see that the radius of convergence of $g(z)$ in (\ref{gz}) is $z^*=1$. 
Then equation (\ref{fugacity}) and the discussion following it shows that a
jamming transition occurs at the critical density 
\be
\varrho_c=\frac{1}{b-2}
\label{cp}
\ee
for $b > 2$. 
Although we will focus on the rate (\ref{hoprate}) which models
`attractive interactions' between particles in the ZRP, we also
consider the case of free particles in Appendix~\ref{app_free} for
which $u(n)=n$ \cite{Jain:2001}. In the latter case, each particle is
endowed with an exponential clock that ticks at rate one, but since
the particles are free and act independently, the total hopout rate is 
equal to the number of particles at the site.

\section{Partition function in the canonical ensemble}
\label{partfn}

\subsection{Exact Recursions}

Consider a system of $N$ particles on $L$ sites. If $\tau$ is a
configuration in this system, let $W(\tau)$ denote the stationary
weight of such a configuration. 
Let $Z_{L,N}$ denote the partition function of the EP in this
system. That is to say, 
\[
Z_{L,N} = \sum_\tau W(\tau).
\]
We will first give two different recurrence relations for $Z_{L,N}$. 

Note that any configuration can be written in the form 
\(
0^{k_0} \tau 1 0^{k_N},
\)
where $\tau$ is a configuration in the system with $L-k_0-k_N-1$ sites
and $N-1$ particles. Since we want this representation to be unique,
$\tau_1$ has to be 1. 
Thus,
\begin{align*}
Z_{L,N} &= \sum_{0 \leq k_0+k_N \leq L-N} \sum_\tau \;
W(0^{k_0} \tau 1 0^{k_N}) \\
&= \sum_{0 \leq k_0+k_N \leq L-N} \sum_\tau \;
f(k_0+k_N) W(\tau) \\
&= \sum_{k=0}^{L-N} (k+1) f(k) \; \sum_\tau W(\tau),
\end{align*}
where we set $k =k_0+k_N$ in the last line and the factor of $k+1$
counts for the number of ways one can split $k$ in this way. The sum
over $\tau$ now gives the partition function for a system with $L-k-1$
sites and $N-1$ particles where the first site is occupied. Since the
system is translation-invariant, this gives the formula, 
\be
Z_{L,N}= \sum_{k=0}^{L-N} f(k) (k+1) \frac{N-1}{L-k-1} Z_{L-k-1,N-1}.
\label{pfnrec1}
\ee

Another recurrence relation for the ZRP partition function ${\tilde Z}_{l,n}$
with $l$ sites and 
$n$ particles has been obtained \cite{Chleboun:2010} and is given by
\be
{\tilde Z}_{l,n}= \sum_{k=0}^n f(k) {\tilde Z}_{l-1,n-k}
\label{pfnreczrp}
\ee
with ${\tilde Z}_{0,n}=\delta_{n,0}$ since ${\tilde Z}_{1,n}=f(n)$. 
Since a ZRP can be mapped to EP by regarding the $N$ particles in EP
as $l$ sites in ZRP and $L-N$ holes in EP as $n$ particles in ZRP,
the two partition functions can be related as 
\be
Z_{L,N}= \frac{L}{N}~ {\tilde Z}_{N,L-N}
\label{map}
\ee
The prefactor on the right hand site (RHS) of the above equation
arises due to the fact 
that the mapping described above between ZRP and EP assumes that an
EP configuration begins with an occupied site. The EP configurations
that begin with an empty site are taken care of by the factor $L/N$
using the argument described above. 

Therefore, on using the last two equations, we get 
\be \label{pfnrec2}
Z_{L,N}= \frac{L}{N}\sum_{k=0}^{L-N} f(k) \frac{N-1}{L-k-1} Z_{L-k-1,N-1}.
\ee
We have thus shown that both recurrence relations \eqref{pfnrec1} and
\eqref{pfnrec2} with the initial conditions $Z_{L,0} = Z_{L,L} = 1$
give rise to the same formula. Although we have proved this result, we
have no deeper understanding of this equivalence.


\subsection{Exact Formula for the Partition Function}
It turns out that one can express $Z_{L,N}$ exactly using integer partitions. To state the result, we need some definitions. An integer partition of a positive integer $n$ is a representation of $n$ in terms of other positive integers which sum up to $n$. For convenience, the parts are written in weakly decreasing order. For example $(5,3,3,2,1)$ is a partition of 14. If $\lambda$ is a partition of $n$, we denote this as $\lambda \vdash n$.

Another way of expressing a partition is in the so-called frequency representation,
$1^{a_1} 2^{a_2} \cdots$, where $a_i$ represents the multiplicity of $i$ in the partition. 
This information can be encoded as a vector $\bar a = (a_1,a_2, \dots )$. 
For example, the same partition of 14 above can be written as $1^1 2^1 3^2 4^0 5^1 \equiv (1,1,2,0,1)$ followed by an infinite string of zeros, which we omit.
We will write $\bar a \vdash n$ to mean a partition of $n$ in this notation.

The  number of parts of a partition, denoted by $|\bar a|$, is given by
$\sum_i a_i$. Given a function $f$ defined on the positive integers, we will denote
\[
f(\bar a) = f(1)^{a_1} f(2)^{a_2} \cdots.
\]
In the same vein, let $\bar{a}! = \prod_i a_i!$. Finally, recall that the  Pochhammer symbol or  rising factorial $(m)_n$ 
for nonnegative integer $n$, is given by the product $m(m+1) \cdots (m+n-1)$ if $n$ is positive and by $m$ if $n=0$.

The partition function of the EP can be written as
\be \label{pfexact}
Z_{N+M,N} = (N+M) \sum_{\bar a \vdash M} \frac{\ds (N-|\bar a|+1)_{|\bar a|-1}}
{\ds \bar{a}!} \; f(\bar a),
\ee
where the length of the system is $L = N+M$ and
$(m)_n$ is the  Pochhammer symbol defined after \eqref{fn}.
We will prove this by equating both representations \eqref{pfnrec1} and \eqref{pfnrec2}. Doing so for $Z_{N+M,N}$ shows
\[
\sum_{k=0}^M \frac{(N-1) f(k)}{N+M-k-1} Z_{N+M-k-1,N-1} \left( \frac{M-Nk}{N} \right) =0.
\]
Isolating the $k=0$ term and replacing $N-1$ by $N$ gives a recurrence
\[
\frac{M}{N+M} Z_{N+M,N} = \sum_{k=1}^M \frac{(N+1)k-M}{N+M-k} f(k) \; Z_{N+M-k,N}.
\]
Define $\hat{Z}_{N+M,N} = \frac{Z_{N+M,N}}{N+M}$ to get a recurrence for $\hat{Z}$'s,
\be \label{zhatrec}
\hat{Z}_{N+M,N} = \sum_{k=1}^M \frac{(N+1)k-M}{M} f(k) \; \hat{Z}_{N+M-k,N}.
\ee
We will now prove the formula for $\hat{Z}_{N+M,N}$ equivalent to \eqref{pfexact} by induction on $M$. When $M=1$, there is a single term in the sum corresponding to $\bar{a}=(1,0,\dots)$. Thus $\hat{Z}_{N+1,N} = f(1)$. This is correct since there is a single vacancy and a factor of $f(1)$ for the particle preceding it.

Now, we assume that \eqref{pfexact} is true for the number of vacancies being any of $1,\dots,M-1$. Using \eqref{zhatrec} and the induction assumption, we can write
\begin{equation}
\hat{Z}_{M+N,N} = \sum_{k=1}^M \frac{(N+1)k-M}{N+M-k} f(k) \; \sum_{\bar{a} \vdash M-k}
\frac{\ds (N-|\bar a|+1)_{|\bar a|-1}}
{\ds \bar{a}!} \; f(\bar a).
\end{equation}
Notice that each term in the above equation contains the factor $f(k) f(\bar{a})$ where 
$a \vdash M-k$. We can thus replace $\bar{a}$ in the sum by $\overline{a'}$, where 
$\overline{a'} = \bar{a} \oplus (k)$. Then  $f(k) f(\bar{a})$ can be replaced by $f(\overline{a'})$. Therefore, the sum above can be reinterpreted as a sum over partitions of $M$. We have to compute the coefficient of $f(\overline{a'})$ in such a term. 

Suppose $\overline{a'}$ can be written as $(i_1^{a'_1}, \dots ,i_j^{a'_j})$ where each $a'_k \neq 0$. Since there are $j$ distinct parts in $\overline{a'}$, we can express $\overline{a'} = (i_k) \oplus \overline{a'}_k$, where $\overline{a'}_k = (i_1^{a'_1}, \cdots, i_k^{a'_k-1}, \cdots i_j^{a'_j})$ for $k=1,\dots,j$. There are thus, exactly $j$ terms that contribute to the partition $\overline{a'}$. Note that
\[
|\overline{a'}_k| = |\overline{a'}| -1, \quad f(\overline{a'}) = f(\overline{a'}_k)f(i_k) \quad \text{ and } 
\quad \overline{a'}! = \overline{a'}_k! \; a'_{i_k}.
\]
The terms contributing to $\overline{a'}$ are
\[
\begin{split}
& \sum_{k=1}^j \frac{(N+1)i_k-M}{M} f(i_k) \frac{\ds (N-|\overline{a'}_k|+1)_{|\overline{a'}_k|-1}}
{\ds \overline{a'}_k!} f(\overline{a'}_k)\\
= & \sum_{k=1}^j \frac{(N+1)i_k-M}{M} \frac{\ds (N-|\overline{a'}|+2)_{|\overline{a'}|-2}}
{\overline{a'}!} \; a'_{i_k} \; f(\overline{a'}) \\
= & \frac{\ds (N-|\overline{a'}|+2)_{|\overline{a'}|-2}}{\overline{a'}!} \; f(\overline{a'}) 
\sum_{k=1}^j \frac{(N+1)i_k a'_{i_k} -M a'_{i_k} }{M} \\
= & \frac{\ds (N-|\overline{a'}|+2)_{|\overline{a'}|-2}}{\overline{a'}!} \; f(\overline{a'}) (N+1-|\overline{a'}|) \\
= & \frac{\ds (N-|\overline{a'}|+1)_{|\overline{a'}|-1}}{\overline{a'}!} \; f(\overline{a'}).
\end{split}
\]
This is precisely what we wanted to show.


\section{Two-point correlation function in canonical ensemble} 
\label{twocorrfn}


\subsection{Exact formula for finite system}

We wish to calculate the two-point connected correlation function
\be
C(r)=\langle n_i n_{i+r} \rangle -\rho^2 ~,~r > 0
\label{corrgenl}
\ee
in a system of $L$ sites with $N$ particles. Let us consider a set of
configurations in which the $r$ sites from $i$ to $i+r-1$ contain $k$
holes. Then the contribution to the  
correlation function $\langle n_i n_{i+r} 
\rangle$ comes from only those configurations in which both the $i$th and
$(i+r)$th site are occupied. Using the mapping between EP and ZRP described in 
Sec.~\ref{model} and summing over all the particle configurations in front of
the $i$th and $(i+r)$th particle, we get   
\bea
\langle n_i n_{i+r} \rangle 
&=& \sum_{k=k_{min}}^{k_{max}} \frac{{\tilde
    Z}_{r-k,k}~{\tilde Z}_{N-r+k,L-N-k}}{{Z}_{L,N}} \\
&=& \rho \sum_{k=k_{min}}^{k_{max}} \frac{{\tilde
    Z}_{r-k,k}~{\tilde Z}_{N-r+k,L-N-k}}{{\tilde Z}_{N,L-N}} 
\label{corrsim}
\eea
where we have used (\ref{map}) to arrive at the last
expression. As the total number of particles is conserved, the maximum 
number of particles in the first cluster can be $N-1$. In other words,
$r-k \leq N-1$ which gives $k_{min}=\max(0,r-N+1)$, as the lower limit
$k_{min}$ can not be below zero. Also the  
local conservation in the first cluster with $r$ sites requires that $k 
\leq r-1$. Thus we find that $k_{max}=\min(L-N,r-1)$ since $k_{max}$
can not exceed the total number of holes in the system.

\subsection{Exact expression for infinitely large system}

\begin{figure}
\centering
\includegraphics[width=1\textwidth,angle=0]{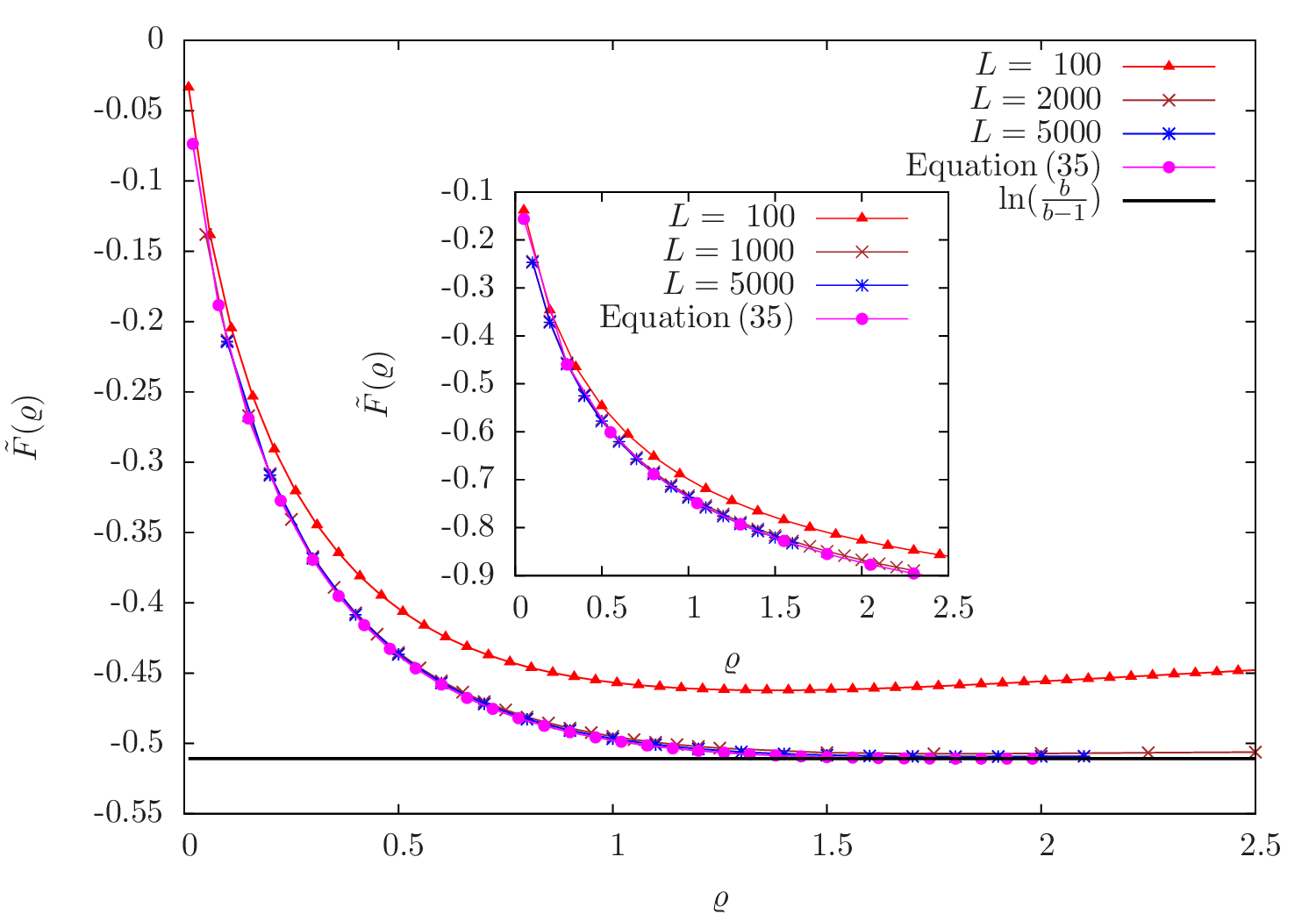}
\caption{Free energy $\tilde{F}(\varrho)$ as a function of density
 $\varrho$ for $b=3/2$ (inset) and $b=5/2$ (main) for different
 system sizes. The data for finite-sized systems is obtained by
 numerically 
 solving the recursion equation (\ref{pfnreczrp}) and is compared with the
 result (\ref{freeE}) for infinitely large system.}
\label{fig_freeenergy}
\end{figure}

It is evident from (\ref{corrsim}) that the partition function at all
densities is required to evaluate the correlation 
function. However, barring some special cases that are discussed in 
Appendix~\ref{app_b0} and ~\ref{app_free}, it does not seem possible to calculate the exact partition
function ${\tilde Z}_{l,n}$ for all densities. In the 
following subsections, we will calculate the two-point correlation  
function in the thermodynamic limit as the problem is analytically
tractable in this limit. For $L \to \infty$ and finite 
$r$, we first note that the limits in the sum appearing in
(\ref{corrsim}) simplify to 
$k_{min}=0$ and $k_{max}=r-1$.  Furthermore, inspired by equilibrium
statistical mechanics, we conjecture that 
there exists a `free energy' ${\tilde F}(\varrho)$ defined as 
\be
{\tilde F}(\varrho) =-\lim_{l \to \infty} \frac{\ln {\tilde
    Z}_{l,n}}{l} 
\label{zrpenergy}
\ee
For the hop rate (\ref{hoprate}), using the recursion equation
(\ref{pfnreczrp}), we calculated the partition function ${\tilde
  Z}_{l,n}$ as a function of density for various system sizes. Figure
\ref{fig_freeenergy} shows that the scaled 
logarithmic partition function indeed approaches a limiting function
with increasing system size.

Thus for large $L$, using (\ref{zrpenergy}), we can write \cite{Huang:2000}
\be
\ln \left( \frac{{\tilde Z}_{N-r+k,L-N-k}}{{\tilde Z}_{N,L-N}} \right)
=k \mu - (r-k) P
\label{ratio}
\ee
where the chemical potential $\mu$ and the pressure $P$ are given by 
\begin{subequations}
\bea
\mu &=& \frac{\partial {(l {\tilde F})}}{\partial n}\Big |_l={\tilde
  F}'(\varrho) \\
P &=& -\frac{\partial {(l {\tilde F})}}{\partial l}\Big |_n=-{\tilde
  F}(\varrho)+\varrho {\tilde F}'(\varrho) 
\eea
\label{thermo}
\end{subequations}
and the prime stands for derivative with respect to $\varrho$. 
Using (\ref{ratio}) in the expression (\ref{corrsim}) for correlation
function $\langle n_i n_{i+r} \rangle$ and the boundary condition ${\tilde Z}_{0,n}=\delta_{n,0}$ (refer the discussion after (\ref{pfnreczrp})), we get
\bea
\langle n_i n_{i+r} \rangle &=& \rho ~e^{-r P} \sum_{k=0}^{r} {\tilde
  Z}_{r-k,k}~e^{k (\mu+P)} ~,~r \geq 0\\
&=& \rho ~e^{r \mu} \sum_{k=0}^{r} {\tilde
  Z}_{k,r-k}~e^{-k (\mu+P)} \label{genfncorr1}\\
&=& \rho ~e^{r \mu} \sum_{k=0}^{r} \frac{k}{r}
~{Z}_{r,k}~e^{-k (\mu+P)} \\
&=&  -\frac{\rho}{r} ~e^{r \mu} 
\frac{d}{d(\mu+P)}  \sum_{k=0}^{r} {Z}_{r,k}~e^{-k  (\mu+P)}
\eea
Thus the correlation function is related to the 
{\it grand canonical} partition function of the EP with
$r$ sites, which is not known.

However as explained in Sec.~\ref{model}, the grand canonical
partition function for ZRP is known. 
We therefore define the generating function of the correlation
function as $G(y) =\sum_{r=0}^\infty y^r C(r)$ which, on using
(\ref{genfncorr1}), works out to be 
\bea
G(y) &=& \rho \sum_{l=0}^\infty (y e^{-P})^l ~ \sum_{n=0}^\infty
{\tilde Z}_{l,n} (y e^{\mu})^n -\frac{\rho^2}{1-y} \\
&=& \frac{\rho}{1-y e^{-P} g(y z)}-\frac{\rho^2}{1-y}
\label{genfncorr2}
\eea
where $z=e^{\mu}$. 
Furthermore, we recall that the equation of state in grandcanonical
ensemble is given by \cite{Huang:2000}  
\be
P l=\ln({\tilde {\cal Z}}_l(z))= l \ln g(z)
\label{state}
\ee
which thus gives $e^{-P}=1/g(z)$. Thus we arrive at our main result, namely 
\be
G(y)=\frac{\rho ~g(z)}{g(z)-y~ g(y z)}-\frac{\rho^2}{1-y}
\label{Gmain}
\ee
where the fugacity $z(\rho)$ is determined by (\ref{fugacity}). The
correlation function is then given by 
\bea 
C(r) &=& \frac{1}{r!} ~\frac{d^r G(y)}{d y^r}\bigg|_{y=0} \label{deriv}\\
&=& \oint_C \frac{dy}{2 \pi i} ~\frac{G(y)}{y^{r+1}}
\label{residue}
\eea
where the integral in the last expression is along the closed curve
$C$ around the origin \cite{Mathews:1969}. We check that $C(0)=\rho
(1-\rho)$ is obtained 
from the above expression. The behavior at $r \to \infty$ is obtained by taking
the limit $y \to 1$. Expanding (\ref{Gmain}) close to $y=1$ and using
(\ref{fugacity}), we see that $G(y)$ (and hence $C(r)$) vanishes as $r
\to \infty$. 

Before proceeding further, we note that due to (\ref{thermo}) and
(\ref{state}), the free energy can be written as   
\be
{\tilde F}(\varrho)=\varrho ~ \mu - \ln g(z)
\label{freeE}
\ee
This expression is also plotted in Fig.~\ref{fig_freeenergy} for hop
rate (\ref{hoprate}) with $b=3/2$ and $5/2$, and we
see that it matches well with the results for large
systems. We note that ${\tilde F}(\varrho)$ is a decreasing  
  function of the density for $b < 2$, but it saturates to $-\ln
  g(1)$ at high density for $b > 2$.

\section{Correlation function for hop rate (\ref{hoprate})}
\label{specific}

\begin{figure}
\centering
\includegraphics[width=1\textwidth,angle=0]{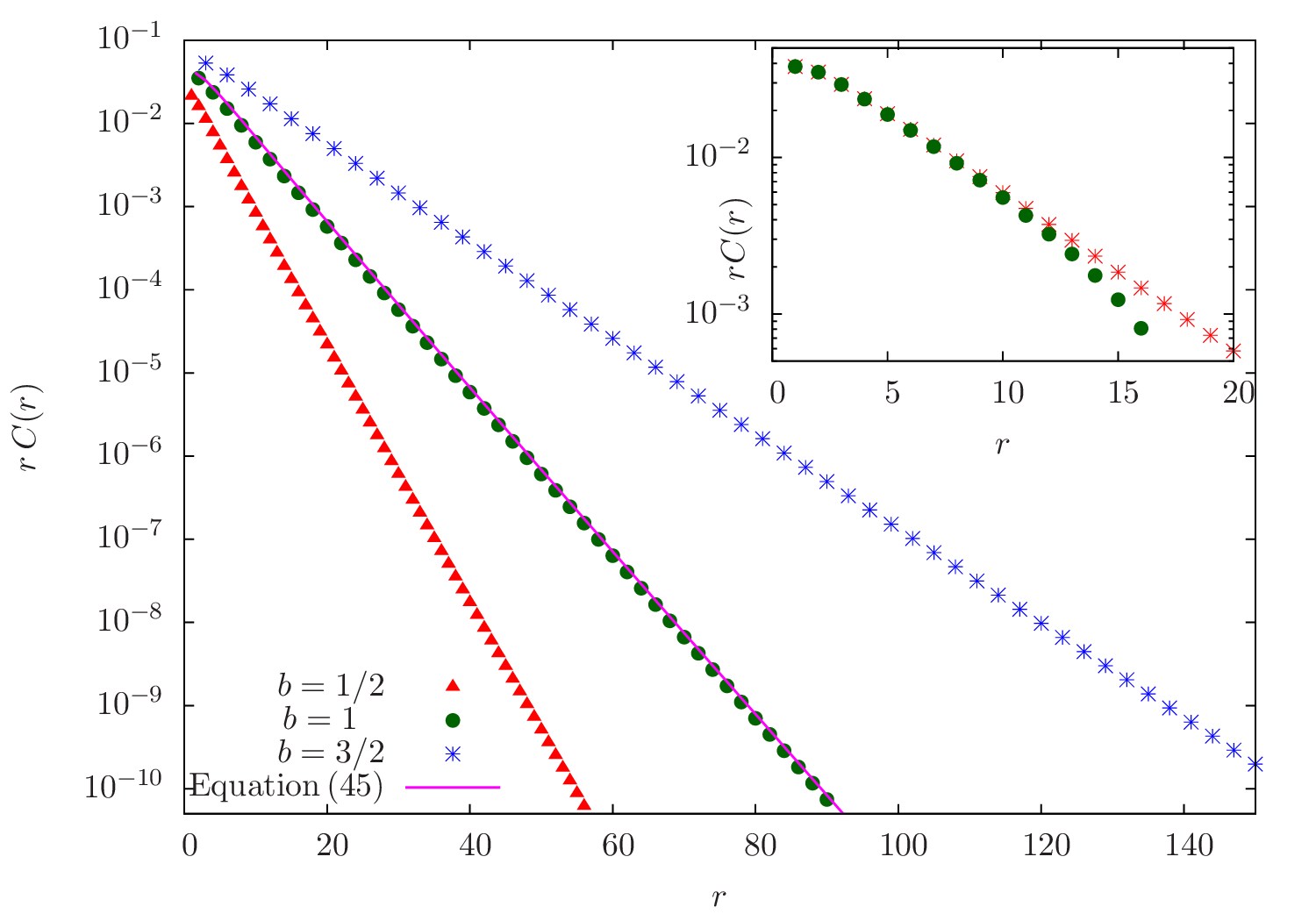}
\caption{Decay of the correlation function in the laminar phase at
 $\rho=0.4$ for various $b < 2$ in the infinite system. The
 analytical result (\ref{b1final}) for 
 $b=1$ is also shown. The inset compares the correlation 
 function for $b=1$ obtained using (\ref{corrsim}) for $L=10^4$ and
 (\ref{Gmain}) for infinite system.}  
\label{fig_laminar}
\end{figure}

\begin{figure}
\centering
\includegraphics[width=1\textwidth,angle=0]{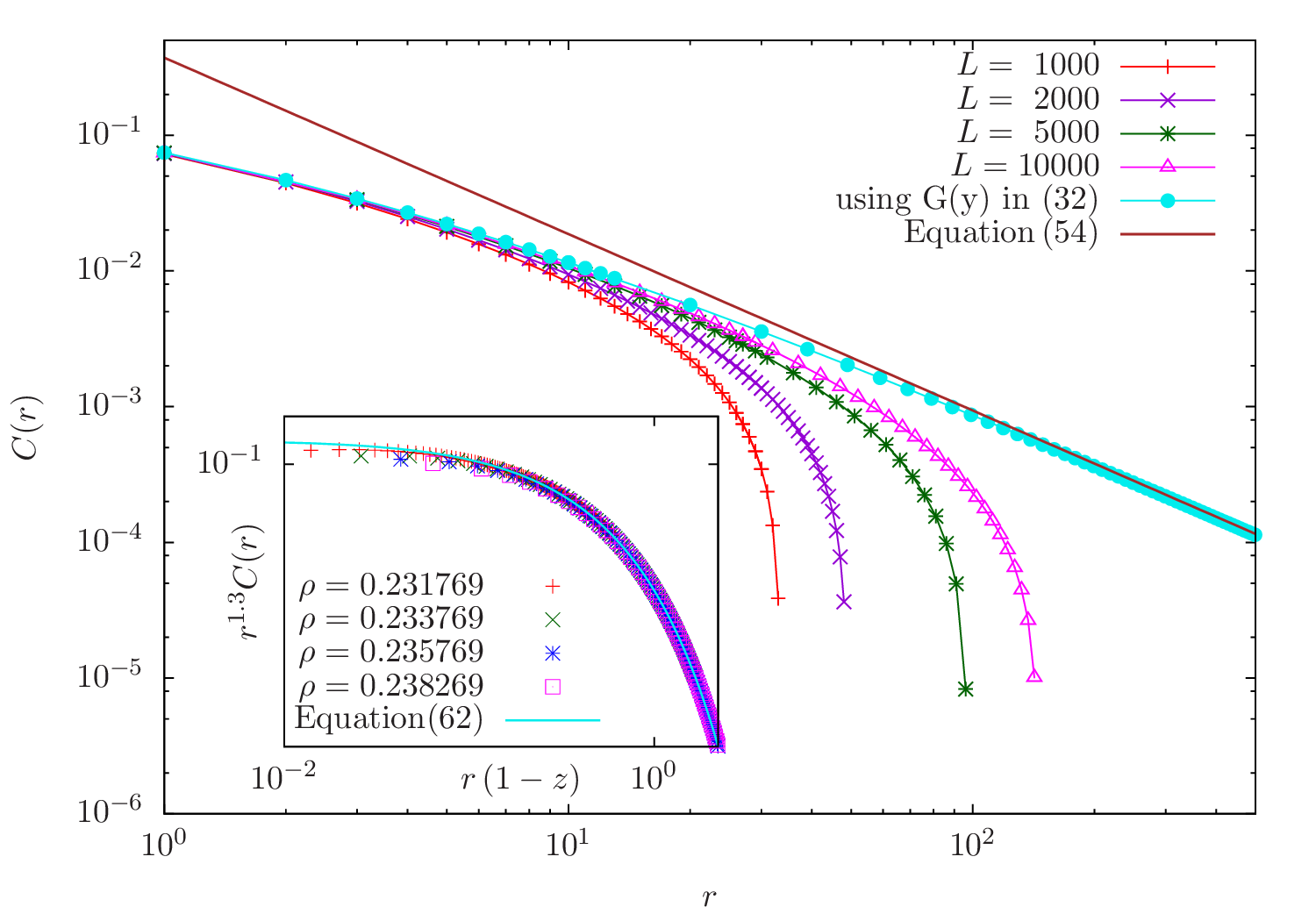} 
\caption{Decay of spatial correlation function with distance at the
 critical density for $b=3.3$. The data for finite-sized systems is
 obtained by numerically solving (\ref{pfnreczrp}) and
 (\ref{corrsim}), while the result in the thermodynamic limit is 
 obtained using (\ref{Gmain}). The analytical result (\ref{invert})
 valid for large inter-particle distances is also shown. The inset
 shows the data collapse 
 of the correlation function for different densities close to the 
 critical point in the laminar phase for 
 infinite system using (\ref{scalingform}).} 
\label{corr3pt3}
\end{figure}

We now apply the general result (\ref{Gmain}) for the generating
function of the correlation
function to the choice (\ref{hoprate}) of the hop rates. The
correlation function can be easily obtained numerically from
(\ref{Gmain}) for an infinitely large system, and these results are 
shown along with those obtained using the exact result (\ref{corrsim}) for a
finite system in Figs.~\ref{fig_laminar}, \ref{corr2pt5} and
\ref{corr3pt3}, and we see that the latter approaches the result
obtained from (\ref{Gmain}) with increasing system size. In the
following subsections, we obtain analytical results for $C(r)$ using
(\ref{Gmain}). 

\subsection{Laminar phase: $0 < b < 2$}

When $b=0$, we obtain the well known TASEP \cite{Spitzer:1970} on a
ring for which the steady state 
is known exactly. This case is discussed briefly in 
Appendix~\ref{app_b0} using (\ref{Gmain}). For $b=1$, the generating
function $g(z)$ given by (\ref{gz}) takes a particularly simple
form: 
\be
g(z)=-\frac{\ln (1-z)}{z}
\ee
Therefore, from (\ref{Gmain}), we get the generating function of the
correlation function as 
\be
G(y)=\frac{\rho \ln(1-z)}{\ln(y_0-1)-\ln(y_0-y)}-\frac{\rho^2}{1-y}
\ee
where $y_0=1/z > 1$. The density-fugacity relation (\ref{fugacity}) is
given by  
\be
\rho= -\frac{1-z}{z}~\ln(1-z)~,~ z < 1
\label{rhozb1}
\ee
To calculate the correlation function, we consider the following integral
in the complex-$y$ plane along a closed contour $C'$ wrapped around the branch cut at $y_0$ which consists of a large circle of radius $R$ about the origin and a small circle of radius $\epsilon$ about $y_0$:
\be
I_1=\frac{1}{2 \pi i}~\oint_{C'} \frac{dy}{y^{r+1}}~ \frac{\rho
  \ln(1-z)}{\ln(y_0-1)-\ln(y_0-y)}
\label{I1cont}
\ee
As the integrand has a simple pole at $y=1$ and poles of order $r+1$ at
$y=0$, due to (\ref{residue}), the residue at these poles immediately
gives $C(r)$. It is easy to check that the contribution from the
integrals over the large and the small circle vanishes when $R \to \infty$ and $\epsilon \to 0$. 
Since the integrand in (\ref{I1cont}) also has a branch cut singularity at
$y=y_0$, we finally obtain  
\be
C(r)= \frac{1}{2 \pi i} \left(\int_{AB} \frac{dy}{y^{r+1}}~ \frac{\rho
  \ln(1-z)}{\ln(y_0-1)-\ln(y_0-y)}+\int_{B'A'} \frac{dy}{y^{r+1}}~ \frac{\rho
  \ln(1-z)}{\ln(y_0-1)-\ln(y_0-y)} \right)
\ee
where $y-y_0=x$ along $AB$ and $y-y_0=x e^{i 2 \pi}$ along
$B'A'$. Since the correlation function is real, writing $-x=x e^{-i
  \pi}$, we get 
\bea
C(r) &=& \frac{\rho \ln(1-z)}{2 \pi i} \int_0^\infty
\frac{dx}{(y_0+x)^{r+1}} ~\left( \frac{1}{\ln(y_0-1)-\ln x+i \pi}-
\frac{1}{\ln(y_0-1)-\ln x-i \pi}\right) \\
&=& \frac{-\rho \ln(1-z) (1-z)}{y_0^r} \int_0^\infty \frac{dx}{(1+x (1-z))^{r+1}}~
\frac{1}{(\ln x)^2+\pi^2}
\eea
We are not able to perform the above integral exactly. But an
approximate expression can be found for large $r$ as follows:
\bea
C(r) &\approx& \frac{-\rho \ln(1-z) (1-z)}{y_0^r} \int_0^\infty
dx~\frac{e^{-r x (1-z)}}{(\ln x)^2+\pi^2} \\
&\approx& \frac{-\rho \ln(1-z) (1-z)}{y_0^r} \int_0^{\frac{1}{r
    (1-z)}} \frac{dx}{(\ln x)^2+\pi^2} \\
&\approx& \frac{\rho |\ln(1-z)|}{r}~ \frac{e^{-r |\ln z|}}{[\ln (r (1-z))]^2+\pi^2}
\label{b1final}
\eea
where the last expression is obtained after an integration by parts
and the fugacity is determined in terms of density from
(\ref{rhozb1}). The last result is plotted against that obtained by
solving (\ref{Gmain}) numerically, and we see an excellent agreement. 
Like the $b=1$ case, in general for $0 < b < 2$, the correlation function
shows an  exponential decay (with power law correction), as can be seen
in Fig.~\ref{fig_laminar}.


\subsection{At the critical density: $b > 2$}

We now calculate the correlation function at the
critical density (\ref{cp}) using (\ref{Gmain}).  
At the critical density $\varrho_c$, as the fugacity $z=1$, we get
\be
G(y)=\frac{\rho_c ~g(1)}{g(1)-y~ g(y)}-\frac{\rho_c^2}{1-y}
\ee
We first consider the case when $b$ is not an integer. 
For large $r$, we can expand $g(y)$ given by (\ref{gz}) about $y=1$. 
Using equation (15.3.6) of \cite{Abramowitz:1964}, we obtain 
\be
g(y)= g(1)-s g'(1)+ \frac{s^2}{2!} g''(1)+...+\frac{(-s)^n}{n!} g^{(n)}(1)+ \alpha s^{b-1}+{\cal
  O}(s^{b}) 
\ee
where $s=1-y$. Here we have retained analytic terms in the  
Taylor series expansion up to $n$th order where $n$ is the integer
part of $b-1$ and the leading non-analytic term.  
In the above expression, $\alpha=b \pi \csc(b \pi)$ and $g(1)=b/(b-1)$.
Then we have 
\bea
G(s) &=& \int_0^\infty dr~e^{-s r} C(r) \\
&=& \frac{\rho_c}{\frac{s}{\rho_c}+c_2 s^2+...+c_n s^n-\frac{\alpha
    s^{b-1}}{g(1)}}-\frac{\rho_c^2}{s}  \\
&=& \frac{\rho_c^2}{s}~ \left[\frac{\alpha \rho_c
  s^{b-2}}{g(1)}-\rho_c (c_2 s+...+c_n s^{n-1})  \right]
\eea
where the coefficients $c_i$ are writeable in terms of the
derivatives of $g(z)$ evaluated at one. Since $G(s)$ has a branch cut
singularity at $s=0$, its inverse Laplace transform is given by
\cite{Mathews:1969} 
\bea
C(r) &=& \frac{1}{2 \pi i} \int_{-i \infty}^{i \infty} ds~ e^{s r} ~G(s)\\
&=& -\frac{\rho_c^3}{2 \pi i} \int_{-i \infty}^{i \infty} ds~
e^{s r} ~\left(c_2+c_3 s+...+c_n s^{n-2}- \frac{\alpha s^{b-3}}{g(1)}
\right)
\label{laplace}
\eea
An integral similar to above also appears in the calculation of the
canonical partition function of the ZRP \cite{Evans:2006} and we can use
those results here. In the above expression, the first integral  is 
$\delta(r)$ and all the 
integrals (barring the last one) are the derivatives of the delta
function. Therefore for large $r$, these integrals vanish, and we are
left with
\bea
C(r) = \frac{\alpha \rho_c^3}{g(1)} \int_{-i \infty}^{i \infty}
\frac{ds}{2 \pi i}~e^{s
  r} ~s^{b-3}
\eea
The above integral can be obtained from the integral $I_2$ calculated
in the Appendix~\ref{app_cont} by setting $c=t=0$, and we obtain 
\be
C(r)= \frac{\rho_c^2 ~\Gamma(b-1)}{r^{b-2}}
\label{invert}
\ee
This result is compared against that obtained using (\ref{Gmain}),
and we see an excellent match at large $r$.  

When $b$ is an integer, as before, we expand $g(y)$ about $y=1$ and 
using (15.3.11) of \cite{Abramowitz:1964}), we obtain 
\be
g(1-s)= g(1)-s g'(1)+ \frac{s^2}{2!} g''(1)+...+\frac{(-s)^n}{n!}
g^{(n)}(1)+\beta s^{b-1} \ln s 
\ee
where $\beta=(-1)^b b$. Following the same steps as described above,
we get 
\bea
C(r) &=& \frac{\beta \rho_c^3}{g(1)} \int_{-i \infty}^{i \infty}
\frac{ds}{2 \pi i}~e^{s r} ~s^{b-3} \ln s \\
&=& \frac{\rho_c^3 \beta}{g(1) r^{b-2}}~\int_{-i \infty}^{i \infty}
\frac{ds}{2 \pi i} ~ e^s s^{b-3} (\ln s- \ln r) \\
&=& \frac{\rho_c^3 \beta}{g(1) r^{b-2}}~\int_{-i \infty}^{i \infty}
\frac{ds}{2 \pi i} ~ e^s s^{b-3} \ln s
\eea
where we have used that $b$ is an integer to arrive at the
last equation. As the above integrand has a branch cut at $s=0$, proceeding in a manner similar to that described in Appendix~\ref{app_cont} with
$c=t=0$, we find the above integral  
to be
$(-1)^{b-2}~\Gamma(b-2)$ which shows that (\ref{invert}) is valid for
integer $b$ as well.


\subsection{Above the critical density: $b > 2$}

We now consider the behavior of the correlation function in the laminar
phase at a density close to the critical point. Since the fugacity is
below one here, we write $t=1-z$ and expand (\ref{fugacity}) about
$z=1$ to find the relationship between $t$ and $\rho$. We find that 
\be
{\frac{1}{\rho}=}
\begin{cases}
\frac{1}{\rho_c}- \frac{\alpha (b-1) g'(1)}{g^2(1)}~t^{b-2} ~,~2 < b < 3 \\ 
\frac{1}{\rho_c}+ \frac{g'(1)}{g(1)} 
\left(\frac{g'(1)}{g(1)}-\frac{g''(1)}{g'(1)} -1\right) t ~,~b > 3 
\end{cases}
\label{bg2}
\ee
The next order corrections to the above expression can also be worked
out, and turn out to be of the order $t$ for $2 < b < 3$, $t^{b-2}$ for $3
< b < 4$ and $t^2$ for $b > 4$.

For large distances and densities close to the critical density, we
now expand the generating function $G(y,z)$ in (\ref{Gmain}) about $y=1$
and $z=1$. For $b > 3$, on using (\ref{bg2}), we obtain
\be
G(s,t)= \frac{\alpha \rho^3}{g(1)}~ \frac{(s+t)^{b-1}-t^{b-1}}{s^2}
\ee
where, as before, $s=1-y$ and we have dropped the analytic terms as
they do not contribute to $C(r,z)$ for the same reasons as described in
the last subsection. We then have 
\be
C(r,z) = \frac{\alpha \rho^3}{g(1)}~\int_{c-i \infty}^{c+i \infty}
\frac{ds}{2 \pi i}~e^{s r} ~\frac{(s+t)^{b-1}-t^{b-1}}{s^2} 
\label{lam_int}
\ee
where $c$ is a positive real number. The above integral is calculated
in Appendix~\ref{app_cont}, and we find that in the limit $t=1-z \to 0, r
\to \infty$ with $r t$ finite, the correlation function is of the
following scaling form
\be
C(r,z)= r^{2-b}~{\cal H}(r (1-z))
\label{scalingform}
\ee
where the scaling function 
\be
{\cal H}(x)= (b-2) \Gamma(b-1) \rho_c^2 ~e^{-x} \left[ (x+b-1) e^x
  E_{b-1}(x)-1 \right] 
\label{scalingfn}
\ee
is a decreasing function of $x$. In the above expression,
$E_n(x)=\int_1^\infty dt ~e^{-x t} t^{-n}$ is
the exponential integral. By carrying out a calculation similar to
above, it can be checked that the results (\ref{scalingform}) and
(\ref{scalingfn}) hold for $2 < b <3$ and integer $b$ as well. The
inset of Fig.~\ref{corr3pt3} shows the data collapse for the correlation
function for various densities close to the critical point and the
scaling function. 

Using the asymptotic properties of the
exponential integral $E_n(x)$ \cite{Abramowitz:1964}, we find that the
scaling function ${\cal H}(x) \stackrel{x \to 0}{\sim} \Gamma(b-1)
\rho_c^2$ which thus 
reproduces the result at the critical point obtained in the last
subsection. At large $x$, as the scaling function ${\cal H}(x)  \stackrel{x \to
  \infty }{\sim} \Gamma(b) (b-2)  \rho_c^2 ~e^{-x}/x^2$, the
correlation function decays exponentially fast with inter-particle
distance $r$. This analysis yields the correlation
length defined by $x= r/\xi$ to be 
\be
\xi \sim (1-z)^{-1} \sim (\rho-\rho_c)^{-\nu}
\label{xidefn}
\ee
which, by virtue, of (\ref{bg2}) gives $\nu=1/(b-2)$ for $b < 3$ and $1$
for $b > 3$.

\section{Discussion}
\label{disc}

In this article, we studied an exclusion process on a ring in which a
particle hops to a right empty neighbour with a rate that depends on the
number of empty sites in front of it. Although we assumed that the
hops are totally asymmetric, the results obtained here 
hold for the general case also in which a particle may hop to either 
left or right empty neighbour with nonzero rate. 
This is because the general exclusion model maps to a ZRP whose 
partition function is independent of the bias in the hop rates
\cite{Evans:2005}. Then our exact equation (\ref{corrsim}) for the
correlation function, which holds for any bias in the hop rates,
gives the same solution as obtained in the previous sections. 

Although most of the results for the ZRP and hence the exclusion
process have been obtained in the grand 
canonical ensemble \cite{Evans:2005}, some studies in the canonical
ensemble have also been carried out
\cite{Evans:2006,Kanai:2007,Evans:2008,Chleboun:2010}. In 
particular, an expression for the partition function ${\tilde
  Z}_{l,n}$ in the canonical 
ensemble at and in the vicinity of the critical point has been
calculated for finite systems \cite{Evans:2006}, and it has been shown
that for the weight $f(m)$ with the same asymptotic behavior as
(\ref{gz}), ${\tilde Z}_{l,n}$ depends exponentially on system size for 
$\varrho < \varrho_c$, but sublinearly on $l$ for $\varrho \geq
\varrho_c$. This implies that the free energy (\ref{zrpenergy})
changes with the density $\varrho$ in the homogeneous phase but
becomes a constant equal to 
$-\ln g(1)$ (which is chosen to be zero in \cite{Evans:2006}) for all
$\varrho \geq \varrho_c$, as seen here in 
Fig.~\ref{fig_freeenergy}. Due to the latter property, our analysis
can not be carried over to the jammed phase. However since we are
mainly concerned with 
critical exponents here, it suffices to consider the system in the
infinite size limit.

For an infinitely large system, we have derived an exact expression
(\ref{Gmain}) for the generating function of the steady state two-point
correlation function in the canonical ensemble. This result was
applied to the hop rate (\ref{hoprate}) for $\rho \geq \rho_c$ to find
the relevant critical exponents.  
Interestingly, we 
find that at the critical point, the exponent characterising the power
law decay of the 
two-point correlation function changes continuously with the parameter
$b$ in the hop rate (\ref{hoprate}). Equilibrium systems in two
dimensions  that show  
continuously varying exponents at the critical point are known 
\cite{Baxter:1982}, and their behavior is understood in terms of
conformal field theories with central charge one \cite{Cardy:1987}. We
do not know if the behavior found here has any such deeper significance. 
The correlation length exponent 
$\nu$ in (\ref{xidefn}) also changes continuously for $2 < b < 3$, whereas
it is constant for $b > 3$. This scaling for the correlation length
has been obtained in a previous work \cite{Kafri:2003} as well. In addition,
we have also derived the scaling function for the correlation function in
the high density phase here. 
The case of $b=1$, where the system is in laminar phase for all
densities, has been considered in \cite{Basu:2010a}, but an 
explicit expression for the correlation function was not provided.

\begin{figure}
\centering
\includegraphics[width=1\textwidth,angle=0]{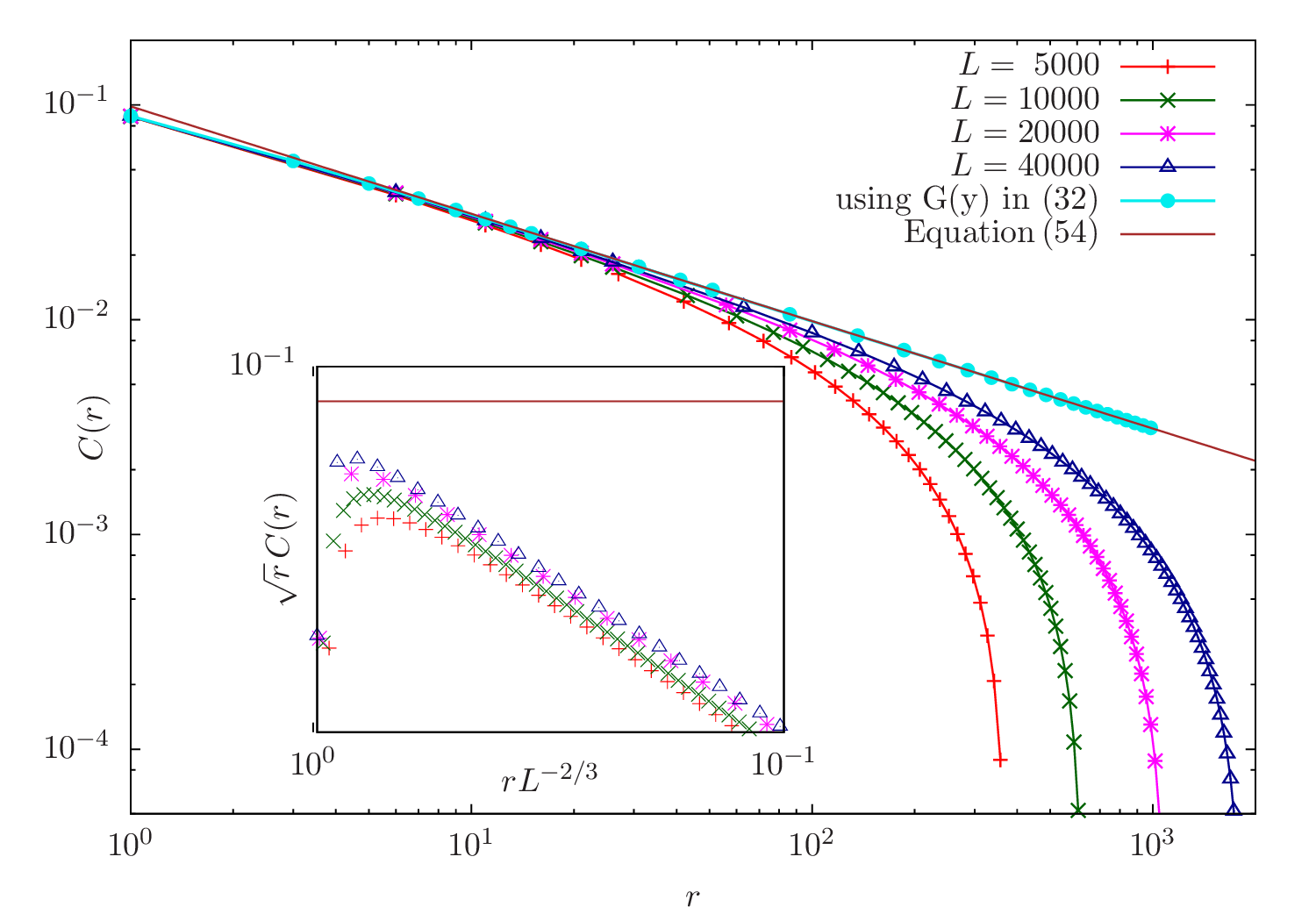}
\caption{Decay of spatial correlation function with distance at the
 critical density for $b=5/2$. The data for finite-sized systems is
 obtained by numerically solving (\ref{pfnreczrp}) and
 (\ref{corrsim}), while the result in the thermodynamic limit is 
 obtained using (\ref{Gmain}). The analytical result (\ref{invert})
 valid for large inter-particle distances is also shown. The inset
 shows the data collapse 
 of $C(r,L)$ for different system sizes using
 (\ref{finiteL}).}  
\label{corr2pt5}
\end{figure}

From the numerical data shown in
Figs.~\ref{corr2pt5} and \ref{corr3pt3} at the critical point, we 
note that the finite 
size effects set in early on. For example, in Fig.~\ref{corr2pt5} for
$b=5/2$ and a system size $L=10^4$, a power law is seen for about a 
decade only. This makes a numerical determination of the correlation 
function exponent difficult. Here we have given an expression
(\ref{Gmain}) for the generating function of the two-point
correlation function for an infinite system which can easily generate
several decades of data. For a finite system with $L$ sites, we
expect the correlation function to be of the following scaling form: 
\be
C(r,L)=\frac{1}{r^{b-2}} ~{\cal F} (r L^{-z})
\label{finiteL}
\ee
where the scaling function ${\cal F}(x)$ is a constant for $x \ll 1$
and decays for $x \gg 1$. In the ZRP, the average mass
cluster at the 
critical point scales as $l^{1/(b-1)}, b < 3$ and $\sqrt{l}, b >
3$ \cite{Evans:2006}. If we make
the reasonable assumption that at the critical density, there is a
single length scale in the system under consideration and is set by
the typical headway, we expect $z=1/(b-1)$ for $b < 3$. This
expectation is consistent with the data shown in the inset of
Fig.~\ref{corr2pt5} for $b=5/2$ where we see that the data collapse
gets better with increasing $L$.  

Acknowledgements: The authors thank S. N. Majumdar and P. K. Mohanty
for helpful discussions, and M. R. Evans for bringing  \cite{Kanai:2007}  to our attention. 


\appendix

\section{Simple exclusion process}
\label{app_b0}

For $u(n)=1~,~n > 0$, as all configurations are equally 
likely \cite{Liggett:1985}, the steady state partition function is
given by $Z_{L,N}={L \choose N}$. The
two-point correlation function $C(r)={L-2 \choose N-2}/{L \choose N}=N
(N-1)/(L (L-1))~,~r > 0$ vanishes in the limit $L \to \infty$. It can
be easily checked that (\ref{corrsim}) also gives this result. 
For large systems, the free energy defined in (\ref{zrpenergy}) works
out to be 
\be
{\tilde F}(\varrho) = (1+\varrho) \ln(1+\varrho)-\varrho \ln \varrho
\ee
which is an increasing function of the density $\varrho$. 
Furthermore, since $f(m)=1$, we have $g(z)=1/(1-z)$ and therefore
\be
G(y)=\rho \frac{1-(1-\rho) y}{1-y}-\frac{\rho^2}{1-y}
\ee
which immediately yields $C(r)=0~,~r > 0$, as expected in the
thermodynamic limit. 

\section{Free particle case}
\label{app_free}

When the particles jump independently (in 
the ZRP picture), the hop out rate is proportional to the number of
particles at the site, $u(n)=n$ . Therefore from (\ref{product}), the ZRP partition function is easily seen to be 
\[  
{\tilde Z_{l,n}}= \frac{l^n}{n!}.
\]
Using this in (\ref{corrsim}), we obtain the exact expression for the two-point
correlation function as 
\be
\langle n_i n_{i+r} \rangle=\frac{\rho}{N^{L-N}} \sum_{k=k_{min}}^{k_{max}}
\frac{(r-k)^k}{k!} (N-r+k)^{L-N-k} ~\frac{(L-N)!}{(L-N-k)!}
\ee
In the thermodynamic limit, the above expression gives 
\be
\langle n_i n_{i+r} \rangle= \rho \sum_{k=0}^{r-1} \frac{(r-k)^k}{k!}~ \varrho^k e^{-(r-k) \varrho}
\label{freeG2}
\ee
For this case, we have $g(z)=e^{z}$ and $z=\varrho$. As a result, (\ref{Gmain}) gives 
\be
G(y)=\frac{\rho}{1-y e^{z (y-1)}}-\frac{\rho^2}{1-y}
\label{freeG}
\ee
It can be checked that the correlation function in (\ref{freeG2})
matches that obtained from the series expansion of (\ref{freeG}).  

To obtain an explicit expression for the correlation function $C(r)$,
we use the Euler-Maclaurin formula given by \cite{Boas:1971}
\bea
\sum_{k=0}^r f(k) &\approx& \int_0^r dx~f(x)+\frac{1}{2}
(f(0)+f(r))-\int_0^r dx~f'(x) \sum_{j=1}^\infty \frac{\sin (2 j \pi
  x)}{\pi j} \\
&=& \int_0^r dx~f(x)+\frac{f(0)}{2}+2 \sum_{j=1}^\infty
\int_0^r dx~\cos(2 j \pi x) f(x) \\
&=& \int_0^r dx~f(x)+\frac{f(0)}{2}+2 \sum_{j=1}^\infty \textrm{Re} \left[\int_0^r
  dx~e^{i 2 j \pi x} f(x) \right]
\eea
where $f(k)$ is the summand in (\ref{freeG2}). Our main task is 
to calculate the integral on the RHS of the last equation which can be
carried out using the saddle point method for large $r$. We find that 
\be
\int_0^r dx~e^{i 2 j \pi x} f(x) \approx \frac{e^{r (x_0-\varrho)}}{1+x_0}
\ee
where $x_0$ is the solution of the saddle point equation 
\be
\varrho-x_0+\ln(\varrho/x_0)+i 2 \pi j=0
\ee
Writing $x_0=\varrho \alpha e^{i \theta}$, we find that $\alpha$ and
$\theta$ obey the following equations:
\begin{subequations}
\bea
\varrho &=& \frac{2 \pi j-\theta}{\tan \theta} + \ln \left( \frac{2 \pi
  j-\theta}{\varrho \sin \theta} \right) \label{theta} \\
\alpha &=& \frac{2 \pi j-\theta}{\varrho \sin \theta} 
\eea
\end{subequations}
For $j=0$, the saddle point $x_0=\varrho$ which immediately gives 
\be
C(r)=2 \rho \sum_{j=1}^\infty \int_0^r dx~\cos (2 j \pi x) f(x)
\ee
where the summand is given by
\be
e^{r (\frac{2 \pi j-\theta}{\tan
      \theta}-\varrho)} ~
  \frac{\cos (r (2 \pi j-\theta)) (1+\frac{2 \pi j-\theta}{\tan
      \theta})+ \sin(r (2 \pi j-\theta)) (2 \pi j-\theta)}{(1+\frac{2
      \pi j-\theta}{\tan \theta})^2+(2 \pi j-\theta)^2} 
\ee
Since the contribution of the successive
terms in the sum decreases with increasing $j$, we estimate only 
the $j=1$ term here. Also, numerical analysis of (\ref{theta}) shows that $\theta$ increases with $j$ and therefore we work within small-$\theta$ approximation. These considerations finally yield 
\be
\theta=\frac{2 \pi}{W(\varrho e^{1+\varrho})}
\ee
where $W$ is the Lambert function that satisfies $W(z) e^{W(z)}=z$
\cite{Corless:1996}, and 
\bea
C(r) &=& 2 \rho e^{-r (\frac{1}{\rho}-\frac{2 \pi}{\theta})} ~
  \frac{\cos (r \theta) (\frac{2 \pi}{\theta})- \sin(r \theta) (2 \pi -\theta)}{(\frac{2 \pi}{\theta})^2+(2 \pi -\theta)^2}  \\
  &\approx& \rho e^{-r (\frac{1}{\rho}-\frac{2 \pi}{\theta})} ~
  \frac{\theta \cos (r \theta)}{\pi}
\eea
which is an oscillatory function with decaying amplitude.

\section{Evaluation of the integral (\ref{lam_int})}
\label{app_cont}

Consider the following integral:
\be
I_2=\frac{1}{2 \pi i} ~\oint_{C'} ds~ e^{s r}~
\frac{(s+t)^{b-1}-t^{b-1}}{s^2}~,~t \geq 0
\label{I2cont}
\ee
where the contour $C'$ around the branch cut at $-t$ includes the Bromwich contour along the line $s=c$, $c$ being real and nonnegative.
The residue
from the second order pole at $s=0$ gives $I_2=(b-1) t^{b-2}$. The
integral along the large semicircle with radius $R$  decays
exponentially fast with increasing $R$, and the one along the small
semicircle with radius $\epsilon$ is proportional to  $\epsilon^{b-2}$
and therefore vanishes as $\epsilon \to 0$. Thus we get  
\be
I_2=\frac{1}{2 \pi i} ~\left( \int_{c-i \infty}^{c+i \infty}+
\int_{AB}+\int_{B'A'}  ds~ e^{s
  r}~ \frac{(s+t)^{b-1}-t^{b-1}}{s^2} \right)
\ee
Since $s=-t+x e^{\pm i \pi}$ along the upper
(lower) branch $AB (B'A')$, we get  
\bea
&&\int_{c-i \infty}^{c+i \infty} \frac{ds}{2 \pi i}~ e^{s
  r}~ \frac{(s+t)^{b-1}-t^{b-1}}{s^2} \\
&=& \frac{\sin(b \pi)}{\pi}~e^{-t r} \int_0^\infty dx~e^{-x r}
\frac{x^{b-1}}{(x+t)^2} + (b-1) t^{b-2}\\
&=& \frac{\sin(b \pi)}{\pi}~\Gamma(b-1) e^{-t r}~ \frac{(r
  t+b-1) E_{b-1}(t r)e^{t r} -1}{r^{b-2}}   
+ (b-1) t^{b-2}
\eea
For $t=c=0$, using that $e^x E_{b-1}(x) \stackrel{x \to 0}{\sim}
(b-2)^{-1}+{\cal O}(x^{b-2})$ \cite{Abramowitz:1964}, the above
integral reduces to 
\be
\int_{-i \infty}^{i \infty} \frac{ds}{2 \pi i}~ e^{s
  r}~ s^{b-3}=\frac{\sin(b \pi)}{\pi}~\frac{\Gamma(b-2)}{r^{b-2}}
\label{intc0t0}
\ee


\clearpage

\clearpage

\clearpage

\clearpage

%

\end{document}